# WT-CFormer: High-Performance Web Traffic Anomaly Detection Based on Spatiotemporal Analysis

Yundi He[1]    Runhua Shi[1]*    Boyan Wang[2]

ydhe@ncepu.edu.cn    rhshi@ncepu.edu.cn    boyanwang@nju.edu.cn

[1]School of Control and Computer Engineering, North China Electric Power University, Beijing 102206, China

[2]School of Intelligence Science and Technology, Nanjing University, Suzhou 215163, China

*Corresponding author

## Abstract

Web traffic (WT) refers to time-series data that captures the volume of data transmitted to and from a web server during a user's visit to a website. However, web traffic has different distributions coming from various sources as well as the imbalance between normal and abnormal categories, it is difficult to accurately and efficiently identify abnormal web traffic. Deep neural network approaches for web traffic anomaly detection have achieved cutting-edge classification performance. In order to achieve high-performance spatiotemporal detection of network attacks, we innovatively design WT-CFormer, which integrates Transformer and CNN, effectively capturing the temporal and spatial characteristics. We conduct a large numbr of experiments to evaluate the method we proposed. The results show that WT-CFormer has the highest performance, obtaining a recall as high as 96.79%, a precision of 97.35%, an F1 score of 97.07%, and an accuracy of 99.43%, which is 7.09%,1.15%, 4.77%, and 0.83% better than the state-of-the-art method, followed by C-LSTM, CTGA, random forest, and KNN algorithms. In addition, we find that the classification performance of WT-CFormer with only 50 training epochs outperforms C-LSTM with 500 training epochs, which greatly improves the convergence performance. Finally, we perform ablation experiments to demonstrate the necessity of each component within WT-CFormer.

## 1. Introduction

As the internet technology advances rapidly, computer networks are increasingly integrated into various aspects of daily life. In these interactions, web servers handle extensive data exchanges, delivering essential resources and information to intended recipients [1]. By 2023, worldwide Internet traffic is projected to reach 167 zettabytes each year [2]. However, as the size of computer networks increases, it provides opportunities for a wide range of attackers to exploit, leading to a variety of malicious attacks based on web traffic. Fig.1 shows some typical types of network attacks and their mapping to web traffic anomaly types. For instance, denial-of-service (DoS) attacks occur when an attacker generates a high volume of requests in a short timeframe, causing a significant surge in web traffic. This overloads the server's resources, preventing it from delivering normal services to legitimate users and potentially causing severe disruptions to web operations [3]. Therefore, in order to prevent malicious attacks from causing huge losses, it is very necessary to do active detection of abnormal traffic on web services.

The work in this paper is dedicated to detecting anomalous traffic on the Web server side. It is a univariate time series classification problem [4]. It consists of a specific window of signals generated by the primary sensors, and then a classifier is used to classify the extracted differential features to identify anomalous traffic activity. The challenge of identifying unusual web traffic patterns is that the traffic characteristics vary significantly based on the type of service and user connection patterns, resulting in highly irregular distributions of web traffic patterns [5], classified into the following three types: 1) point anomalies, which involve specific data samples identified as anomalous and can typically be easily recognized through visual inspection; 2) contextual anomalies, where an anomaly is present within the context of a specific data instance; and 3) collective anomalies, in which a group of related data instances collectively exhibit anomalous behavior, even if the individual instances appear normal. [6] found that contextual and collective anomalies are difficult to detect because they tend to exist within traffic and have patterns similar to normal signals, making it difficult to distinguish abnormal traffic.

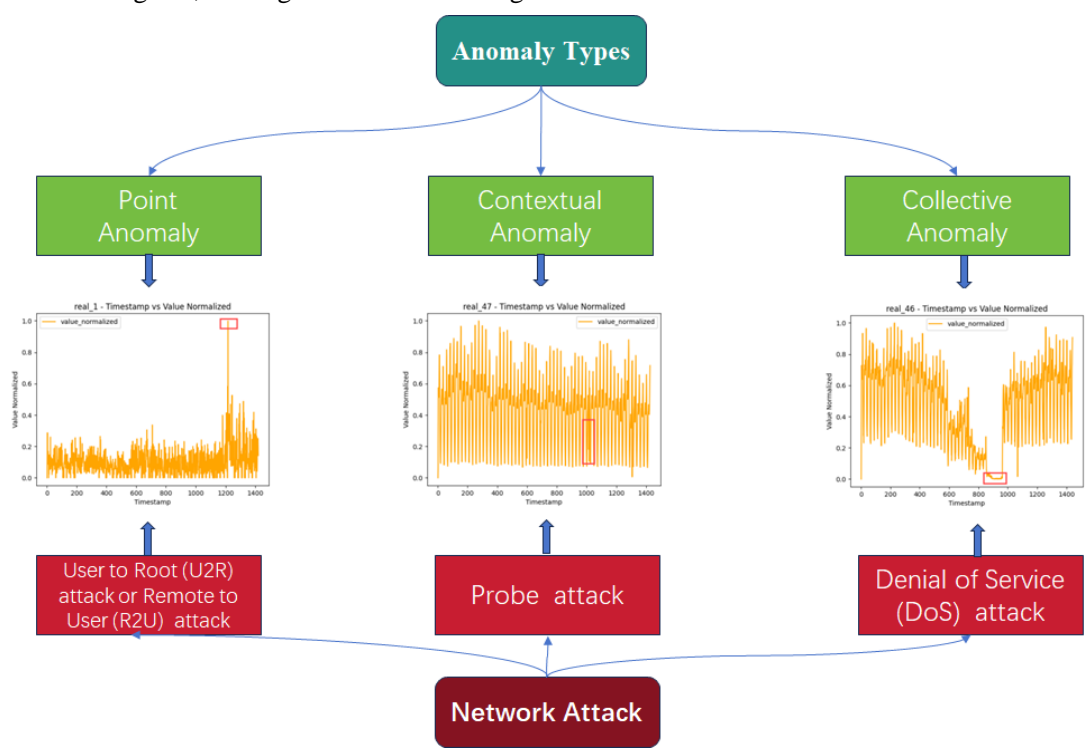

**Fig.1 Mapping figure of anomaly types and attack types.**

Traditionally, web traffic anomaly detection is mainly based on statistical methods. By analyzing the statistical distribution of the characteristics of raw traffic, such as the distribution of the mean and variance of the traffic, thresholds or rules are set to identify anomalous traffic [7]. However, due to the dramatic increase in the scale of web traffic, the anomaly detection models of such methods tend to perform poorly in the face of large-scale or complex web traffic. Subsequently, traditional machine learning-based approaches are applied to address this problem on the ground, obtaining decent performance, but requiring manually designed models to capture features of Web traffic [8]. In recent years, deep learning approaches have been applied to traffic prediction and accurately detect the abnormal traffic data [9]. Compared with traditional machine methods, deep learning has the ability to learn features automatically and don't need much expert knowledge, thus it is well suited to capture the complex and hidden features from a lot of raw traffic data over space and time [10], and accurately identify whether the traffic data is a normal or abnormal pattern. For example, C-LSTM,the state of the art method, effectively models the spatial and temporal information in web traffic data [6]. Based on this, [11] proposed CTGA, a hybrid spatiotemporal neural network with attention, to identify anomalous traffic. In sequences, this author evaluated the performance as not as good as C-LSTM although he innovated the use of TCN and BiGRU instead of the LSTM module to capture short-term features and long-term dependencies in the data, respectively, and used a self-attention mechanism to obtain important information in the sequences. Although current deep learning methods have made some progress in traffic anomaly detection tasks, there are still some drawbacks: 1) The web traffic data in real world is complex and imbalanced, which is hard to capture the characteristics of rare categories, thus reducing the anomaly detection rate; 2) Compared with based on the Transformer, the web traffic anomaly detection based on C-LSTM has a significantly slower convergence rate, resulting in low efficiency in model training, more communication rounds, and deficiencies in detection accuracy as well, hindering its application in the analysis of real web traffic. .

To address these problems, we propose a novel web traffic anomaly detection model called WT-CFormer, which allows cybersecurity practitioners to identify anomalous traffic more comprehensively and accurately in complex network environments. Unlike existing deep learning models, WT-CFormer innovatively use the Transformer Encoder to extract temporal features of web traffic. Transformer captures the dependencies between arbitrary positions in the input sequence through the Self-Attention mechanism, which makes Transformer very effective in dealing with the dependencies of long sequences of web traffic data, and at the same time, it does not rely on the output of the previous moment, and can be processed in full parallel to improve the training efficiency. Secondly, the feed-forward network layer of the Transformer network enhances the expressive ability of the model by enabling it to capture deeper feature information through nonlinear transformations. In addition, we incorporate CNN networks to extract the spatial features of web traffic, propose a more sophisticated design for the corresponding optimisation function of the network architecture of WT-CFormer, and make the proposed WT-CFormer more suitable for processing web traffic sequence data. To demonstrate the effectiveness and superiority to WT-CFormer, we conduct extensive experiments on public datasets and evaluate the anomaly detection performance of our model. Considering the performance of all components within the designed WT-CFormer model, we also perform ablation experiments to demonstrate the necessity of each component of our model.

In summary, the contribution of this paper can be summarized as follows:
1) We preprocess the web traffic sequence data with a sliding window technique, which strengthens the features of samples in minority categories.
2) We design WT-Cformer, that makes a more sophisticated fusion design of Transformer Encoder and CNNs to capture spatiotemporal features, achieving timely and accurate detection of web traffic anomalies. To the best of our knowledge, WT-CFormer firstly adopts the Transformer network in web traffic analysis tasks. Due to its powerful parallelization and global perception capabilities, it significantly improves the classification performance.

3) We evaluate the performance of WT-CFormer on a public web traffic dataset through extensive experiments. Compared with the existing methods, WT-CFormer outperforms all of them. Additionally, its training epoch is less than the state-of-art method, meaning that its fast convergence helps to obtain robust models.

The rest of the paper is organized as follows. Section 2 discusses the related work on web traffic anomaly detection. Section 3 details the input and model architecture of the proposed WT-CFormer. Section 4 describes the experimental and evaluation setup. Section 5 presents the experiment results and analysis. In Section 6, our work is summarized.

## 2. Related Works

In this section, we review related work that addresses anomaly detection in web traffic sequences. We categorize these studies into four types: statistically based detection, spatial feature-based detection, temporal feature-based detection, and spatio-temporal feature-based detection.

### A. Detection based on Statistics

Statistical-based anomaly detection for web traffic sequences is a method to identify anomalous behavior by analyzing statistical features in web traffic sequence data. It detects anomalies by comparing the actual observed traffic features with the normal traffic model or by comparing the predicted traffic with the actual traffic, which can detect anomalous traffic and thus identify possible network attacks or anomalous behavior. The detection method proposed in [13] relied on Markov chain modeling. The transfer probability matrix of the Markov chain is created by modeling the normal web traffic. When new traffic enters, the system determines whether there is an anomaly by calculating the likelihood value of the sequence. If the likelihood value deviates from the normal model, it is labeled as an anomaly. The use of higher-order Markov chains can further improve the detection accuracy and reduce the false alarm rate, thus obtaining a high classification performance. [12] introduced a statistical method for detection of malicious network traffic based on Gaussian Mixture Model (GMM). They learned normal communication behavior by modeling the time overhead of the communication process as GMM generation and predicted anomalies by calculating the likelihood probability of a new data point from the learned GMM distribution. It is intuitively understood that the time overhead of normal communication will be similar to the historical pattern, yielding a high likelihood. Anomalous traffic can deviate from these patterns, yielding a low likelihood. These methods have achieved good performance in classifying statistical anomalies in web traffic data. However, their drawback is that they cannot correctly classify anomalous traffic data that has the same distribution as normal traffic data.

## B. Detection based on spatial features

[15] employed CNN for anomaly detection, leveraging their strong performance in image processing to extract spatial features from packet payloads for effective classification. Similarly, [14] developed an innovative online anomaly detection system utilizing software-defined networks, which employs convolution neural networks to directly analyze raw spatial features of network streams, enabling real-time packet extraction and detection. These approaches represent spatial features by transforming complex sequences into images, allowing CNN algorithms to effectively capture spatial information. While these methods demonstrate superior classification performance compared to previous studies, they face a critical limitation: the loss of temporal information during convolution and pooling operations when dealing with time series data.

## C. Detection based on temporal features

[16] proved a method that combines Conditional Variation Autoencoder (CVAE) and LSTM networks to recognize and detect anomalous traffic. CBAM model suggested by [17] is based on bi-directional LSTM networks that predict conditional event probabilities by learning short-term sequential patterns in web traffic to effectively detect multiple remote access attacks. They extract temporal information by using LSTM models on preprocessed data. These methods model the temporal features of sequence data and classify them using prediction-based algorithms. After learning only normal sequences using RNN, the predicted traffic data is compared with the actual traffic data and anomaly detection is performed based on thresholds. The advantage of this method is that it can periodically predict web traffic data and achieve high classification performance. However, if the pattern of web traffic data does not have a specific period, the predicted traffic data cannot correctly classify the actual traffic data.

## D. Detection based on temporal-spatial features

Web traffic sequence detection based on temporal and spatial features is used to identify anomalous behavior by analyzing the temporal and spatial characteristics of web traffic data. This takes into account not only the variation of traffic in time, but also the spatial characteristics of the traffic. By combining temporal and spatial features, anomalous traffic or potential network attacks can be detected more comprehensively. [11] explored a hybrid spatio-temporal neural network model called CTGA, which identifies anomalous traffic in web traffic by automatically extracting the temporal and spatial features of the sequences and using these features. [6] proposed C-LSTM model by using preprocessed sliding window data as input. The spatial features in the traffic window are extracted by convolutional and pooling layers. Then LSTM layer extracts the temporal features. The trained model then performs anomaly detection on the test data using softmax classifier to provide superior performance by modeling temporal and spatial information of web traffic, which provides state-of-the-art results in web traffic anomaly detection.

# 3. Method

In this section, we first introduce the inputs to our proposed WT-CFormer model. Subsequently, we present the architecture of the WT-CFormer model and the optimization process of our method.

## A. The Inputs of WT-CFormer

At first, Transformer [18] network was developed for NLP and later Transformer attempts were used for feature extraction of sequential data. For Web traffic, depending on the user's access to different web services with a large number of interactions with the web server, the web server generates a large number of time-ordered traffic measurements. In this paper, we construct Web traffic time series through data preprocessing methods as input to WT-CFormer. Previous studies have shown that two steps, serialization and normalization, are important in Web traffic data preprocessing. First, the sequence data is processed into fixed length sequence data and features are extracted from the fixed sequence length data. However, recognizing abnormal category features from sequence data is difficult because the proportion of data in abnormal categories is much less than that in normal categories, i.e., there is a very serious data imbalance problem. Therefore, we apply the sliding window algorithm to solve this problem [19], the mathematical expression of which can be represented by Equation (1)(2). It gradually slides over the data by fixing the window length and then performs local feature extraction on each subset of the sequence data (i.e., elements within the window) to identify anomalous behaviors or attack patterns. To test the proposed method, window samples were created by applying the sliding window algorithm and anomalous window samples were labeled. In addition, since the inputs to the WT-CFormer are values between 0 and 1, we normalize these values using the min-max normalization equation (3) to preprocess the traffic values for anomaly detection.

Assuming that there is a time series dataset $\{x_1, x_2, \ldots, x_n\}$ of length $n$ with sliding window size $w$ and sliding step size $s$, the formula for the subsequence generated by the sliding window is as follows.

$$y_i = \{x_i, x_{i+1}, \ldots, x_{i+w-1}\} \quad (1)$$

where $i$ is the starting index of the sliding window, $i$ starts at 1 and slides each time in step $s$. $w$ is the size of the window. $x_i$ is the $i^{th}$ element of the original sequence. $y_i$ is the sequence of data within the window that starts at position $i$.

The total number of sliding windows can be calculated using the following formula:

$$k = \left\lfloor \frac{n-w}{s} \right\rfloor + 1 \quad (2)$$

where $n$ is the total length of the time series. $w$ is the size of the window. $s$ is the size of step. $\lfloor \cdot \rfloor$ denotes downward rounding.

$$x' = \frac{x - x_{min}}{x_{max} - x_{min}} \quad (3)$$

where $x$ is the value of the actual flow data, $x_{min}$ is the minimum value in the actual flow data, $x_{max}$ is the maximum value in the actual flow data, and $x'$ is the normalized value.

We utilize actual traffic data as input to WT-CFormer. We use x to denote the amount of data sent and received by the web server, and then the input sequence can be expressed as $X = [x_1, x_2, ..., x_n]$, where $x_i \in [0,1]$ and $i \in \{1,2,...,n\}$, $n$ is the number of data in the web traffic sequence. The length of the sequence and the batch size are important for the speed and classification accuracy of traffic anomaly detection. In this paper, we set the length of the sequence to 60 and the batch size to 512. In fact, with the help of the Transformer network in temporal feature extraction, compared with the state-of-the-art deep learning models [6], WT-CFormer can use shorter training time to achieve comparable performance, and the superiority of our proposed method will be demonstrated in the following sections.

## B. The Architecture of WT-Cformer

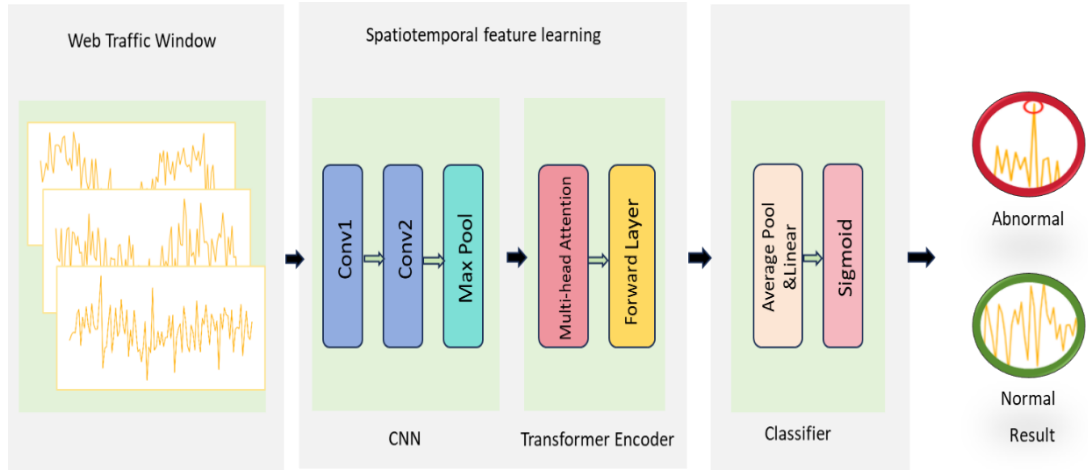

**Fig. 2 The architecture of the proposed WT-CFormer.**

Since web traffic is a typical time series, our proposed WT-CFormer utilizes the Transformer network to extract the temporal features of web traffic, and the CNN network to extract the local properties of the sequence. In this paper, in order to make the proposed model suitable for web traffic, we design a fusion of CNN and Transformer networks.

Fig.2 shows in detail the architecture of the proposed WT-CFormer network, which is composed of CNNs and Transformers connected in series. The spatial features in the flow window are extracted by the Convolutional and Pooling layers. Then the Transformer layer extracts the temporal features. The trained model then performs anomaly detection on the test data using a sigmoid classifier.

First, the CNN layer is composed of two one-dimensional convolutional layers and one maximal pooling layer, which are used to automatically extract higher-level spatial feature sequences of web traffic. These convolution operations utilize a convolution kernel to slide over the sequence and detect features in order. The convolutional layer is followed by an activation function, and unlike the state-of-the-art deep learning model C-LSTM which uses tanh activation function, this paper uses ReLU activation function to enable the CNN to capture complex features in the input sequence. Because, ReLU is simple and has low computational overhead, only the maximum of the input and zero is needed; meanwhile, ReLU remains linear in the positive interval, so it does not have the problem of gradient vanishing when moving away from the origin as the tanh activation function does, and it can backpropagate more efficiently. Therefore, ReLU is used as the activation function after convolutional layer in this paper. Next, the input is given to the Maxpooling layer. The pooling layer reduces the spatial size of the representation to reduce the number of parameters and computational complexity of the network, which also plays a role in preventing overfitting. Maxpooling is a type of pooling that selects the largest number in the partition. Selecting the maximum value in the partition is effective because it indicates the presence of larger activations when features are present. Therefore, maximum pooling has better performance than average pooling for web traffic space feature extraction.

In the Transformer layer, we extract temporal features of web traffic using only the encoder of the Transformer rather than the entire Transformer architecture, this feature makes it easier to understand temporal relationships on large time scales. The encoder block consists of two main sublayers: a multi-head attention layer and a feedforward network. In addition, each sublayer has an Add and Norm module. The output values of the previous pooling layer are used as inputs to the Transformer layer, which performs a linear transformation on the input sequence, see equation (4). After first going through the Multi-head Attention module, the Transformer's self-attention mechanism allows for parallel processing of all positions of the input sequence, whereas the LSTM must process the input sequence sequentially due to its recursive nature. As a result, Transformer is faster in processing long sequences, especially in the training phase, and is better able to capture long-distance dependencies.

In this paper, 8 attention heads are used to focus on different parts of the sequence from different perspectives, and each head performs the attention computation independently, see Equation (5), and then the results are spliced and linearly transformed, see Equation (6), so that the model performs feature extraction in different subspaces to capture richer features. After layer regularization, see Equation (7), it makes the input distribution more stable, which helps to accelerate the training and improve the convergence speed of the model.

$$Q = XW^Q, K = XW^k, V = XW^v \qquad (4)$$

$$head_i = Attention(Q_i, K_i, V_i) = softmax\left(\frac{Q_i K_i^T}{\sqrt{d_k}}\right) V_i \qquad (5)$$

$$MultiHead(Q, K, V) = Concat(head_1, head_2, \ldots, head_h)W^o \qquad (6)$$

$$Attention\ Output = LayerNorm(x + MultiHead(x)) \qquad (7)$$

where Q is the query matrix, K is the key matrix, and V is the value matrix; $W^Q, W^k, W^v$ are the trainable weight matrices; h denotes the number of attention headers, each of which has dimension $d_k = \frac{d_{model}}{h}$; and $W^o$ denotes the output linear transformation matrix.

The output after the attention mechanism is used as an input to the feedforward neural network, which consists of two linear transformations with modified linear unit (ReLU) activation in between. The model setup in this paper maps the input vector x to a 256-dimensional high-dimensional space after one linear transformation, followed by an intermediate high-dimensional space after the ReLU activation function that is mapped back to the original dimensional space by a second linear transformation. The two linear transformations combined with the ReLU nonlinear activation function are capable of extracting and combining higher-level features, which is defined as (8). Layer regularization and residual connectivity are again performed before outputting the feedforward neural network, see equation (9).

$$FFN(x) = ReLU(W_1 x + b_1)W_2 + b_2 \quad (8)$$
$$FFN\ Output = LayerNorm(Attention\ Output + FeedForward(Attention\ Output)) \quad (9)$$

where, $W_1$, $W_2$ are trainable weight matrices and $b_1$, $b_2$ are bias vectors.

As for the classifier, we WT-CFormer model web traffic to be classified by converting the output sequence into detection probabilities. First, it performs average pooling on the output sequence to generate a single vector characterizing the entire sequence. Unlike the state-of-the-art C-LSTM which uses softmax as a classifier, we use Sigmoid as a classification function to map the output values to the range [0, 1] and directly output the anomaly probability values for web traffic detection, see equation (10), which makes the classification more intuitive and efficient.

$$\sigma(z) = \frac{1}{1+e^{-z}} \quad (10)$$

where z is the linear output of the model, and σ(z) is the Sigmoid output (predicted probability) of the model.

In this paper, we use a grid search method to find the optimal parameters of the WT-CFormer model, as shown in Algorithm 1.

| Algorithm 1 | Proposed WT-CFormer Model |
|---|---|
| Require: | Training data (X train, y train) |
| Ensure: | Ensure: Trained detection model |
| 1: | Set Nepochs as number of training epochs |
| 2: | for epoch = 1 to Nepochs do |
| 3: | for each batch in (X train, y train) do |
| 4: | Apply sliding window to input sequence |
| 5: | with window size = 60 |
| 6: | for each sliding window in sequence do |
| 7: | x window ← Get window slice from input |
| 8: | conv output ← CNN(x window) |
| 9: | pooled ← MaxPool(conv output) |
| 10: | transformer output ← TransformerEncoder(pooled) |
| 11: | |
| 12: | detection output ← Classifier(transformer output) |
| 13: | |
| 14: | loss ← LossFunction(detection output, y train) |
| 15: | Backpropagate and update model weights |
| 16: | end for |
| 17: | end for |
| 18: | end for |
| 19: | Save model after training |
| 20: | |

# 4. Experiment and Evaluation

In this section, we conduct extensive experiments to demonstrate the effectiveness and superiority of our proposed method. Our proposed WT-CFormer innovatively utilizes the Transformer network and fuses CNNs to perform spatio-temporal information feature extraction work on web traffic. We then experiment and evaluate WT-CFormer on a public web traffic dataset and compare it with the state-of-the-art methods.

## A. Dataset

In this paper, we use the Yahoo S5 Webscope datasetA1 class from Yahoo, a company that operates a U.S. web portal, as an experimental dataset. This dataset is a collection of traffic measurements generated by real Web services, which are represented in chronological order (captured once an hour and packed in a file). The outliers are manually labeled and the data has relatively high traffic variations compared to other available datasets. The dataset contains 67 files with a total of 94,866 data values, but only 1,669 of these data values are anomalous, just about 0.02%. Thus, traffic anomaly detection is a data imbalance problem, using data with an unusually small percentage of anomalous values, significantly affecting the performance of classification algorithms. After data preprocessing, a total of 90,913 window samples are obtained, of which 8,556 anomalous window samples are used for experimental evaluation. The dataset ratio is divided as training set: test set = 7:3, i.e., the training set contains 63639 windows while the test set contains 27274 windows. %More details can be found in Table 1.

Table 1. The details of web traffic Dataset

| Dataset | #of files | #of data /file | Total data | Abnormal data | Total traffic windows | Abnormal traffic windows | traffic windows (Training set) | traffic windows (Testing set) |
|---|---|---|---|---|---|---|---|---|
| Yahoo S5 Webscope dataset A1 | 67 | About 1500 | 94866 | 1669 | 90913 | 8556 | 63639 | 27274 |

## B. Experiment Details

Our proposed WT-CFormer method has a novel deep network architecture for web traffic feature extraction. Compared to existing deep web traffic anomaly detection methods with C-LSTM as the feature extractor, the proposed WT-CFormer utilizes Transformer, which incorporates improvements in web traffic feature extraction. The implementation of our WT-CFormer model uses the Pytorch framework [20]. We utilize Adam as an optimizer to train the WT-CFormer network, with the learning rate set to 0.001. The batch size is set to 512. Unlike the state-of-the-art C-LSTM model which is trained in 500 epochs, our proposed WT-CFormer model achieves higher performance than C-LSTM in just 50 epochs. This greatly reduces the training time and speeds up the convergence of the model. For the dropout layer, we set the rate to 0.1. In addition, we give the hyperparameter search range and selected values in Table 2. In our experiments, we follow the original settings provided in the paper and perform parameter fine-tuning by traversing all candidate parameters through the grid search method to find the optimal ones.

Table 2. Hyperparameters Selection for WT-CFormer

| Hyperparameters | Search Range | Final |
| --- | --- | --- |
| Output channel of Conv#1 | [2,4,8,16,32,64,128] | 64 |
| Kernel size of Conv#1,2 | [2…16] | 5 |
| Stride of Conv#1,2 | [0,1,2,3,4] | 1 |
| Padding of Conv#1,2 | [0,1,2,3,4] | 2 |
| Output channel of Conv#2 | [16,32, 64,128,256] | 64 |
| Activation Functions | [Tanh, ReLU, ELU, GELU] | ReLU, Tanh |
| Pooling Layer | [Average, Max] | Max, Average |
| Kernel size of Pooling | [0…5] | 5 |
| Stride of Pooling | [0,1,2,3,4] | 2 |
| Input Dimension | [50…100] | 60 |
| Number of heads | [1…10] | 8 |
| Dimension of Forward Layer | [64,128,256,512] | 256 |
| Optimizer | [Adam, Adamax, SGD] | Adam |
| Learning Rate | [0.001…0.01] | 0.001 |
| Training Epochs | [10…500] | 50 |
| Min-batch Size | [16…512] | 512 |
| Number of Encoder Layer | [1…5] | 1 |
| Number of FC Layer | [1…4] | 2 |
| Dropout | [0.1…0.8] | 0.1 |
| Classification Function | [Sigmoid, Softmax] | Sigmoid |

## C. Experiment Setup

To evaluate the performance of our proposed WT-CFormer, we use 70% of the dataset of Yahoo S5 Webscope class A1 as the training set to train the WT-CFormer model, and the trained model is tested with 30% of the test set for validation. In our experiments, the optimal parameters are found by tuning a large number of candidate parameters. In the evaluation phase, due to the time-sensitive nature of the web traffic anomaly detection task, which is different from the previous machine learning settings, we try to vary the epoch number and show the superiority of the proposed WT-CFormer. The training epoch is set in the range of [30,50,100,500]. The parameters of the WT-CFormer model with optimal performance are obtained through extensive experiments.

To further ensure the accuracy of the model results, we use a 10-fold cross-validation experiment where the average of the 10 evaluations is used as the final performance metric of the model. It is compared with several state-of-the-art machine learning methods for web traffic anomaly detection, namely KNN, Random Forest, CTGA, and C-LSTM.The KNN selects the nearest K samples for majority voting or averaging in order to determine the classification results or predicted values. Random Forest improves the accuracy and stability of the model by constructing multiple independent decision trees and voting or averaging the predictions of these trees. CTGA employs CNN to extract spatial features of the time series and applies TCN and Bi-directional GRU (BiGRU) to capture short-term local features and long-term dependencies, respectively, and selects the important features using self-attention with weights. weighted self-attention to select the important temporal information, and finally the classification results are obtained by the fully connected layer and softmax layer. The optimal performance C-LSTM is a combination of CNN to extract local features and LSTM to capture long-term dependencies in the time series for processing complex time-series data and multidimensional features.

In addition, to demonstrate the necessity of all the components inside our designed WT-CFormer model, we perform ablation tests.

### D. Evaluation Metrics

We use the predicted probability of the classifier output for anomalous web traffic identification. In our experiments, we evaluate all the methods using the precision, recall, and F1score of web traffic anomaly detection as defined below (11) (12) (13) (14):

$$Accuracy = \frac{TP+TN}{TP+TN+FP+FN} \quad (11)$$

$$Precision = \frac{TP}{TP+FP} \quad (12)$$

$$Recall = \frac{TP}{TP+FN} \quad (13)$$

$$F1\ Score = 2 \times \frac{Precision \times Recall}{Precision+Recall} \quad (14)$$

Where precision (TP is True Positive and FP is False Positive) is the proportion of samples predicted by the model to be positive cases that are actually positive cases. It reflects the accuracy of the model in predicting positive cases. High accuracy means that the model has high confidence in the detected anomalous traffic (positive cases) and low false positives. Recall (FN for False Negative) is a particularly important metric for anomaly detection and is the proportion of samples that are actually positive examples that are correctly predicted as positive examples by the model. It reflects the sensitivity of the model in identifying positive examples. A high recall means that the model is able to detect more anomalous traffic (positive examples) and has a low miss rate. The F1-score is a reconciled average of precision and recall, which is used to evaluate the model's precision and recall ability together. It strikes a balance between precision and recall and is a composite metric. a high F1-score means that the model strikes a good balance between precision and recall, especially for positive and negative sample imbalances. Whereas accuracy (TN is True Negative) indicates the overall proportion of correct predictions of the model, i.e., the proportion of correct predictions among all samples that include both positive and negative cases, it can be misleading in class-imbalanced datasets, especially when there are fewer abnormal samples in the anomaly detection. Even if the model predicts most normal samples correctly, the accuracy may be high, but the prediction of a small number of abnormal samples may be poor.

## 5. Experiment Results and Analysis

### A. Performance Evaluation of WT-CFormer

In this section, we evaluate the performance of WT-CFormer in Web traffic anomaly detection scenarios by conducting experiments on a publicly available Web traffic dataset and presenting the experimental results and related analysis. Specifically, we train the WT-CFormer model using the dataset of the Yahoo S5 Webscope class A1, which enables us to analyze and evaluate the ability of WT-CFormer to detect anomalous web traffic. As mentioned earlier, WT-CFormer is specifically designed for spatio-temporal feature extraction of web traffic.

In addition, we use 10-fold cross-validation to ensure more reliable results. To fully explore the contribution of WT-CFormer, we utilize the five baseline methods mentioned above for comparison. For the baseline methods, we compare WT-CFormer with KNN, Random Forest, CTGA and C-LSTM. We maintain the parametric criteria of the original papers \citep{6}\citep{11}, and compare the highest performance of these machine learning methods with WT-CFormer. KNN sets the number of neighbors to 7. All points in each neighborhood have equal weights. Random Forest sets the number of trees in the forest to 20. we use the Gini to measure the quality of the segmentation. The Gini coefficient is used as entropy for information gain. CTGA uses CNN for spatial feature extraction, TCN for temporal feature extraction, BiGRU for long and short term feature extraction, and self-attention mechanism. C-LSTM uses 2 layers of convolutional and pooling layers and fuses the LSTM layers to extract the spatio-temporal features, and then performs the binary classification task by using Softmax function. The experimental results found that the accuracy of the proposed WT-CFormer is 99.43%. Compared to above machine learning methods, WT-CFormer has the highest performance followed by C-LSTM, CTGA, Random Forest and KNN algorithms as shown in Fig.3.

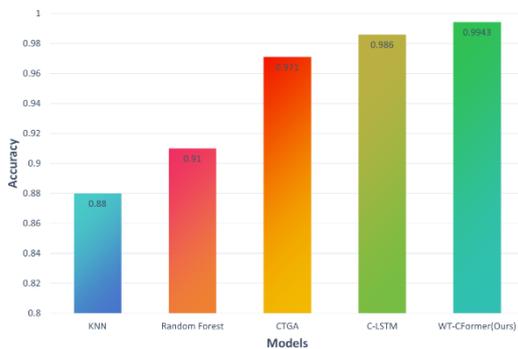

**Fig. 3 Comparing the accuracy of different baseline models**

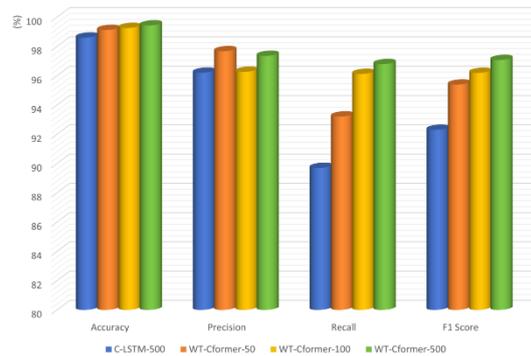

**Fig.4 Performance comparison of C-LSTM and WT-CFormer. (The number represents the training epoch.)**

In the evaluation phase, unlike previous machine learning setups, we vary the epoch count and show the superiority of the proposed WT-CFormer. Note that we use C-LSTN as the baseline method because C-LSTN exhibits better anomaly detection than other machine learning methods. In Table 3, we show the performance results of C-LSTM with WT-CFormer for different training epochs. In general, previous deep web traffic models are trained with a setting of 500, such as the C-LSTM model. However, the larger the training epoch, the higher the training cost of the model and the slower the convergence, because the web traffic anomaly detection task is time-sensitive, which requires shorter training time to make the model converge faster.

Table 3. Performance comparison of different models (%)

| Epoch | 30 | | | | 50 | | | | 100 | | | | 500 | | | |
|---|---|---|---|---|---|---|---|---|---|---|---|---|---|---|---|---|
| Model | Acc | prec | Rec | F1 | Acc | prec | Rec | F1 | Acc | prec | Rec | F1 | Acc | prec | Rec | F1 |
| C-LSTM* | - | - | - | - | - | - | - | - | - | - | - | - | 98.6 | 96.2 | 89.7 | 92.3 |
| WT-CFormer(Ours) | 98.94 | 96.13 | 92.84 | 94.46 | **99.12** | **97.67** | **93.21** | **95.39** | 99.26 | 96.26 | 96.12 | 96.19 | **99.43** | **97.35** | **96.79** | **97.07** |

As shown in Table 3, we demonstrate the performance comparison of different training epochs for the web traffic anomaly detection task. For the evaluation of WT-CFormer, we show the results under four training epoch settings, {30,50,100,500}, and use the state-of-the-art C-LSTM model as the baseline method. Compared to the baseline method, WT-CFormer achieves the best performance in all metrics. For a training epoch of 500, WT-CFormer obtains consistently the best performance for all of them, with a recall of 96.79%, a precision of 97.35%, an F1Score of 97.07%, and an accuracy of 99.43%, respectively, which significantly outperforms the state-of-the-art C-LSTM model. In addition, we can find that the WT-CFormer recognition precision, recall, F1Score and accuracy, improve with the increase of the training epoch, as shown in Fig. 4. For the training epoch set to 50, WT-CFormer still achieves the best performance over the baseline method (Recall 93.21% vs. 89.7%; Precision 97.67% vs. 96.2%; F1 Score 95.39% vs. 92.3%; and Accuracy 99.12% vs. 98.6%), and the training time takes only 139s, which is better than the C-LSTM's training time (769s), which is nearly 6 times shorter than that of C-LSTM, greatly improving the model's learning and convergence speeds, and is especially suitable for traffic anomaly detection tasks with high timeliness requirements. According to our experiments, C-LSTM performs poorly in web traffic anomaly detection, which is attributed to the network architecture choice and design. In terms of feature learning, although the C-LSTM model selects the LSTM network to extract temporal features of web traffic, the performance is still not as good as our choice of Transformer network, and the number of pooling layers of C-LSTM is more than that of WT-CFormer, which reduces the spatial dimensions of the input features and thus inevitably leads to information loss, which proves that the WT CFormer's superiority in feature extraction for web traffic. In terms of classifiers, we choose the sigmoid classifier over the softmax classifier. Compared to sigmoid, softmax complicates the computation and may make the prediction probability of a few classes too low for dealing with web traffic class imbalance data, affecting the model's ability to detect anomalies.

In summary, our proposed WT-CFormer model has obvious effectiveness and superiority and becomes the state-of-the-art web traffic anomaly detection model.

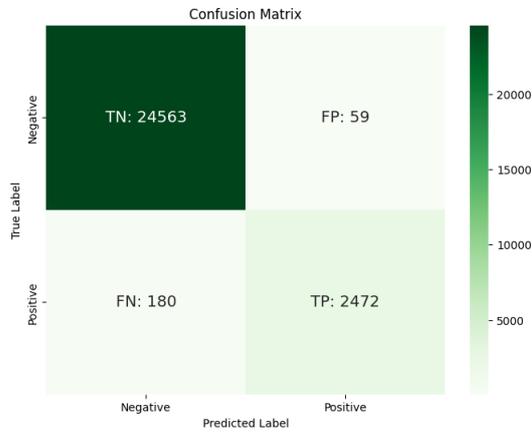 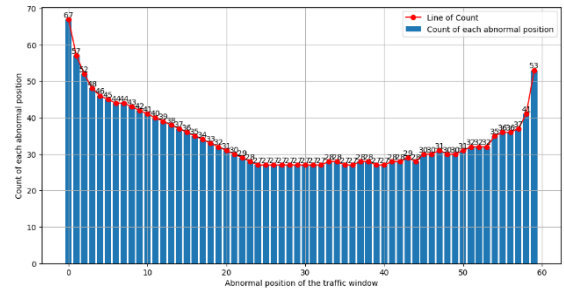

**Fig 5. Confusion Matrix for WT-Cformer.**

**Fig 6. Number of misclassifications for each anomaly value position.**

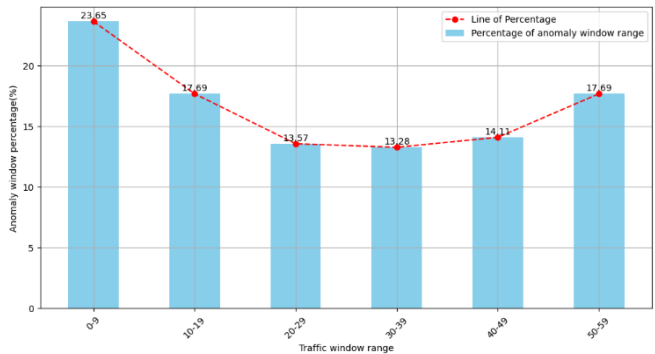 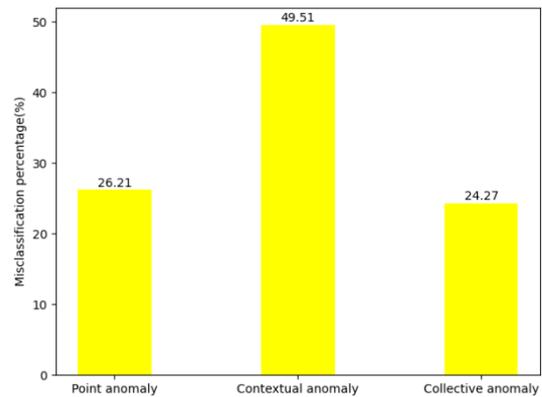

.

**Fig 7. Percentage Distribution of misclassifications for each anomaly value position (Grouped by 10).**

**Fig 8. Statistics of misclassifications for anomaly types.**

In this experiment, we further analyze the misclassified data. Since there are few anomalous traffic windows in anomalous traffic detection, there is an obvious data imbalance problem, so it is very important to accurately classify them. According to the previous analysis, WT-CFormer with the number of training rounds set to 50 performs better than C-LSTM with the number of training rounds set to 500. Therefore, this paper then analyses WT-CFormer with training rounds set to 50.

The experimental dataset is divided into training set and test set in the ratio of 7:3. The confusion matrix data of WT-CFormer is obtained as shown in Fig.5. Among them, the number of traffic windows in which WT-CFormer fails in anomaly classification is 180, and the number of traffic windows in which anomaly detection succeeds is 2472, which substantially improves the classification performance over the previous machine learning methods. In order to further analyse the data of anomaly classification failure, Fig.6 locates the misclassification of anomaly locations for each traffic window, and Fig.7 counts the percentage of the number of misclassified anomaly locations after the traffic window has been divided into six sections. It can be seen that the number of misclassifications is low when the anomalies are located in the middle, and higher when the anomalies are at either end of the traffic window. We also analysed the percentage distribution of anomaly categories not detected by WT-CFormer, as shown in Fig.8. It is clear that it is more difficult to detect contextual anomalies, followed by point anomalies and collective anomalies.

## B. Ablation Experiments of WT-CFormer

In order to study the effect of each component of the WT-CFormer model, we also perform ablation tests on WT-CFormer in the following areas.
(1) CNN: we use only the CNN module to do anomaly detection on web traffic. This method indicates that only the spatial features of web traffic are extracted, while the extraction of temporal features of web traffic is ignored.
(2) Transformer: we discard the learning of spatial features of web traffic and directly use Transformer's multi-head attention mechanism and feed-forward network to efficiently learn the temporal features of web traffic sequences.
(3) WT-CFormer: to illustrate the effectiveness and superiority of WT-CFormer, we fuse CNN and Transformer to learn the spatio-temporal features of web traffic, and learn the information of web traffic more comprehensively to get better anomaly detection performance.

Table 4. Performance comparison of WT-CFormer model internal components (%)

| Method | Accuracy | Precision | Recall | F1score |
|---|---|---|---|---|
| CNN | 97.56 | 96.40 | 77.52 | 85.94 |
| Transformer | 94.03 | 88.16 | 43.95 | 58.65 |
| WT-CFormer（Ours） | **99.12** | **97.67** | **93.21** | **95.39** |

The results of the ablation test are shown in Table 4, and several key observations can be made. First, it is clear that the complete WT-CFormer model outperforms all non-complete models after fusing the CNN and transformer components. Second, of all the model components, the CNN design component provides the most significant performance improvement. Therefore, removing this component leads to the largest decrease in the recall, precision, F1 score, and accuracy of the model. Third, although the Transformer design component achieves a relatively high accuracy rate for normal and anomalous web traffic, a more important metric for the anomaly detection task is the degree of recall of the anomalous traffic, and the Transformer's recall of the anomalous traffic is only 43.95%, with an F1 score of 58.63%, which is a significant improvement over the performance of the WT- CFormer full model performance by almost 50%. Therefore, although the Transformer component can extract the temporal characteristics of web traffic relatively well, it is not suitable to go alone for the task of anomaly detection of web traffic.

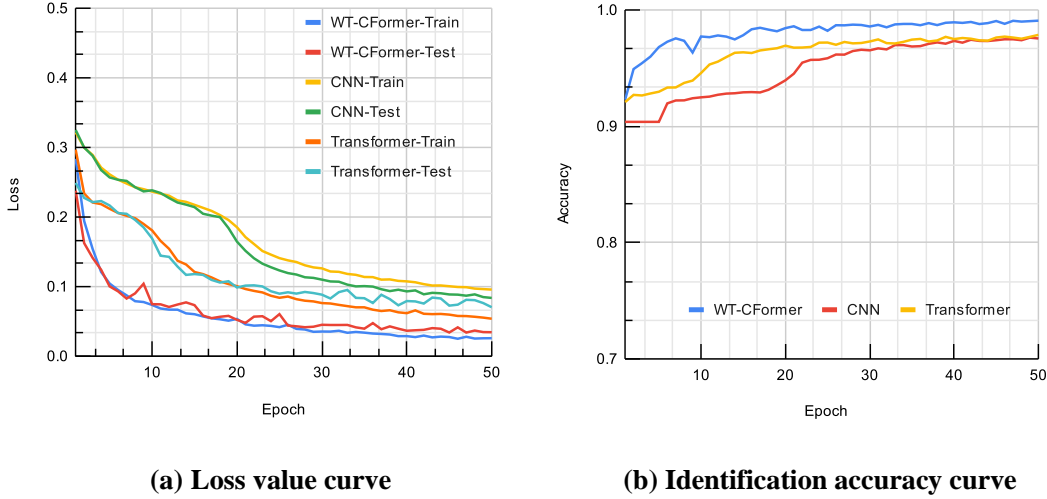

(a) Loss value curve  (b) Identification accuracy curve
**Fig.9 The loss value and Identification accuracy change for WT-CFormer, CNN and Transformer**

In addition, we demonstrate the training process of WT-CFormer on the web traffic anomaly detection task heavy to show the convergence performance of various model models. As shown in Fig. 4, we demonstrate WT-CFormer with a training epoch of 50 as well as CNN and transformer methods with the same training epoch. Fig. 9(a) shows the variation of training and testing losses during training. The WT-CFormer method has lower training and testing loss values than the CNN and transformer methods, and we can find that the testing loss values of our proposed WT-CFormer are almost unchanged after 30 epochs, whereas the losses of the CNN and transformer still maintain a significant decreasing trend after 30 epochs, which means that our proposed WT-CFormer converges faster than CNN and transformer methods. In Fig. 9(b), we show the recognition accuracies of WT-CFormer, CNN and transformer during the training period. Our proposed WT-CFormer obtains a recognition accuracy of more than 99%, while the recognition accuracy of CNN and transformer is about 95%, and the recognition accuracy of WT-CFormer is consistently higher than that of CNN and transformer in each training epoch. The recognition accuracy of WT-CFormer does not change much after 20 epochs After 20 epochs, the recognition accuracy of WT-CFormer does not change much, while the recognition accuracy of CNN and transformer remains stable after 40 epochs, which indicates that WT-CFormer has better convergence than CNN and transformer.

## 6. Conclusion

We proposed a high-performance WT-CFormer architecture for web traffic anomaly detection. We found the optimal model through extensive parametric experiments, including model comparison experiments and ablation experiments. We demonstrated the usefulness and superiority of the proposed model by comparing it with other state-of-the-art machine learning methods as well as with the modules inside the model. It can quickly and accurately detect various anomalies in complex network flows. Our proposed WT-CFormer automatically extracted patterns in network flow data containing spatio-temporal information. It innovatively utilized the Transformer network to extract temporal features of web traffic, and its multi-head attention mechanism captured the dependencies between arbitrary positions in the web traffic sequence as well as parallel computation of multiple attention heads, which greatly accelerated the training speed and provided strong support for the detection of web traffic anomalies with high timeliness requirements. Second, the feed-forward network layer of Transformer network enabled the model to capture the complex features of web traffic through nonlinear transformation, which improved the recognition accuracy of the model. Meanwhile, we designed to deeply integrate the Transformer network into the CNN network and improved important parameters, such as the ReLU activation function and the Sigmoid classifier, etc., which further improved the anomaly detection performance of the WT-CFormer model as a new optimal performance web traffic anomaly detection model. In addition, we found that the performance of WT-CFormer with 50 epochs of training is better than that of previous state-of-the-art models with 500 epochs, which greatly improved the model learning and convergence speed. A large number of experiments in this paper demonstrated the effectiveness and superiority of our proposed WT-CFormer.

However, since the proposed model used a grid search method to find the optimal parameters, especially with many hyperparameters and a wide range of values for each parameter, the experiments were computationally intensive, which led to low efficiency. This problem remains to be optimized in the future. Further research on the fusion design of CNN and Transformer network is also needed to get a better performance web traffic anomaly detection model.

## Acknowledgments

This work was supported in part by National Natural Science Foundation of China (No.62406111, No. 61772001) and Beijing Municipal Natural Science Foundation (No.4242030).